\documentclass[aps,twocolumn,nofootinbib]{revtex4-1}
\usepackage{amssymb}
\usepackage{amsmath}
\usepackage{graphicx}
\usepackage[sort&compress]{natbib}
\bibpunct[, ]{}{}{,}{s}{,}{,}

\usepackage[version=3]{mhchem} 
\begin{document}

\title{Biomolecule surface patterning may enhance membrane association}

\author{Sergey Pogodin, Nigel K. H. Slater$^{\dagger}$ and Vladimir A. Baulin$^{\dag}$}
\affiliation{Departament d'Enginyeria Quimica, Universitat Rovira
i Virgili, Av. dels Paisos Catalans 26, 43007 Tarragona, Spain,
$^\dagger$Department of Chemical
Engineering and Biotechnology, University of Cambridge, Pembroke
Street, Cambridge CB2 3RA, UK and
$^\dag$ICREA, Passeig Lluis Companys 23, 08010 Barcelona, Spain}

\email{vladimir.baulin@urv.cat}

\begin{abstract}
Under dehydration conditions, amphipathic Late Embryogenesis
Abundant (LEA) proteins fold spontaneously from a random
conformation into  $\alpha$-helical structures and this transition
is promoted by the presence of membranes. To gain insight into the
thermodynamics of membrane association we model the resulting
$\alpha$-helical structures as infinite rigid cylinders patterned
with hydrophobic and hydrophilic stripes oriented parallel to
their axis. Statistical thermodynamic calculations using Single
Chain Mean Field (SCMF) theory show that the relative thickness of
the stripes controls the free energy of interaction of the
$\alpha$-helices with a phospholipid bilayer, as does the bilayer
structure and the depth of the equilibrium penetration of the
cylinders into the bilayer. The results may suggest the optimal
thickness of the stripes to mimic the association of such protein
with membranes.
\end{abstract}





\maketitle
Journal link:

\url{http://pubs.acs.org/doi/abs/10.1021/nn204736b}

Certain biological molecules interact with cell membranes to
affect some functional property. Late Embryogenesis Abundant (LEA)
proteins are an example of such molecules. These proteins are
expressed in plant and some animal cells in response to
desiccation conditions and they increase the ability of such cells
to withstand desiccation stress\cite{Covarrubias}. They are
widespread among higher plants but have also been found in some
microorganisms and nematodes exposed to water
deficit\cite{Browne}. LEA proteins may exist in the form of random
coils\cite{Macherel,Wise} and can spontaneously self-assemble into
$\alpha$-helices upon desiccation stress\cite{Tunnacliffe}. Such
an  $\alpha$-helix structure in turn provides an ordered pattern
of alternating hydrophilic and hydrophobic stripes that run along
the axis of the helix\cite{Macherel} which may facilitate the
interaction and insertion of the protein into the phospholipid
membrane. Recently, Popova \emph{et al.}\cite{Popova} have shown that the
fraction of LEA7 protein that takes up the $\alpha$-helix
conformation is increased when it is dried in the presence of
liposomes, indicating that phospholipid membranes associate
closely with LEA proteins and play a critical role in stabilizing
their $\alpha$-helix conformation. LEA proteins are amphiphilic
and when in the $\alpha$-helix conformation can be approximately
described as solid rods with a distinct axial arrangement of
hydrophilic and hydrophobic stripes. An entirely distinct class of
membrane-active peptides, pore-forming peptides\cite{Smart}, also
have surface patterning with a hydrophilic stripe along the
backbone of an $\alpha$-helix. However, the mechanism of
association of these proteins with the phospholipid membrane is
very different. Membrane pores are formed due to the concerted
self-assembly of several peptides on the membrane with a
preferential orientation of peptides perpendicular to the bilayer
plane; the so-called "Barrel and Stave" mechanism\cite{Huang1}. To
allow for such an orientation, the length of the pore-forming
peptides does not usually exceed the thickness of the
bilayer\cite{DesernoPore,Sarkisov}. In contrast, LEA proteins
adopt a more random structure and are much longer\cite{Wise2}. As
a consequence they lie as single molecules parallel to the surface
of the membrane bilayer. The hydrophobic groups on the side of the
backbone control the depth of insertion of the LEA proteins into
the bilayer.

To gain insight into the factors that influence the interaction of
such a rod with a phospholipid bilayer we have calculated the free
energy of association as a function of stripe geometry and
position of the rod relative to the center of the bilayer.

\begin{figure*}
\includegraphics[width=17cm]{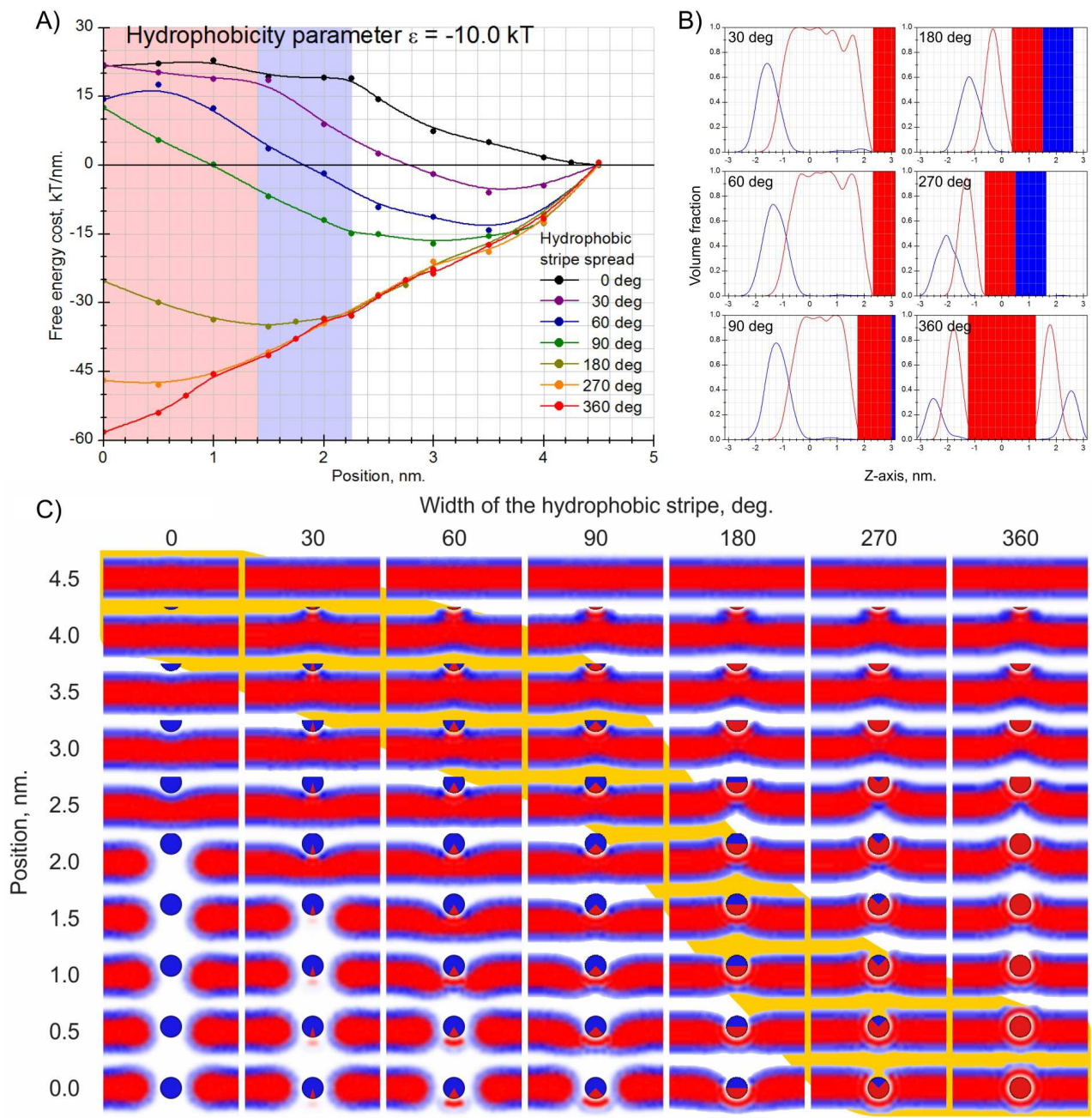}
\caption{Insertion of a hydrophilic cylinder, $\varepsilon=0$,
with a hydrophobic stripe, $\varepsilon=-10$ kT, into phospholipid
bilayer. (a) The free-energy cost of insertion of cylinders with
different width of the hydrophobic stripe (in degrees) as a
function of the center position. (b) The concentration profiles of
tails (red lines) and heads (blue lines) in the central cross
section. Red and blue stripes represent the cylinder with
corresponding stripes. (c) Perturbation of the bilayer (blue:
heads H, red: tails T) upon insertion of the cylinder. The pink
and blue areas at the background show the regions of hydrophobic
core and hydrophilic surface of the unperturbed bilayer. Yellow
band represents the minimum of the free energy (a), which
corresponds to the equilibrium position of the cylinder.}
\label{Results10}
\end{figure*}

In a previous paper\cite{Pogodin3} we showed that Single Chain
Mean Field (SCMF) theory could be used to calculate the
equilibrium structures for insertion of a hydrophobically
patterned rod into a phospholipid bilayer with the rod axis
oriented perpendicular to the layer. These calculations aimed to
determine the effect of patterning upon the puncture of the
bilayer by the rod.  Here, equilibrium structures of association
of a hydrophobically patterned rod with a bilayer are simulated
with the rod axis oriented parallel to the layer. These
calculations aim to mimic the behaviour of moieties such as LEA
proteins that may associate with, but not puncture, bilayers.

\section{Results and discussion}

\begin{figure*}
\includegraphics[width=17cm]{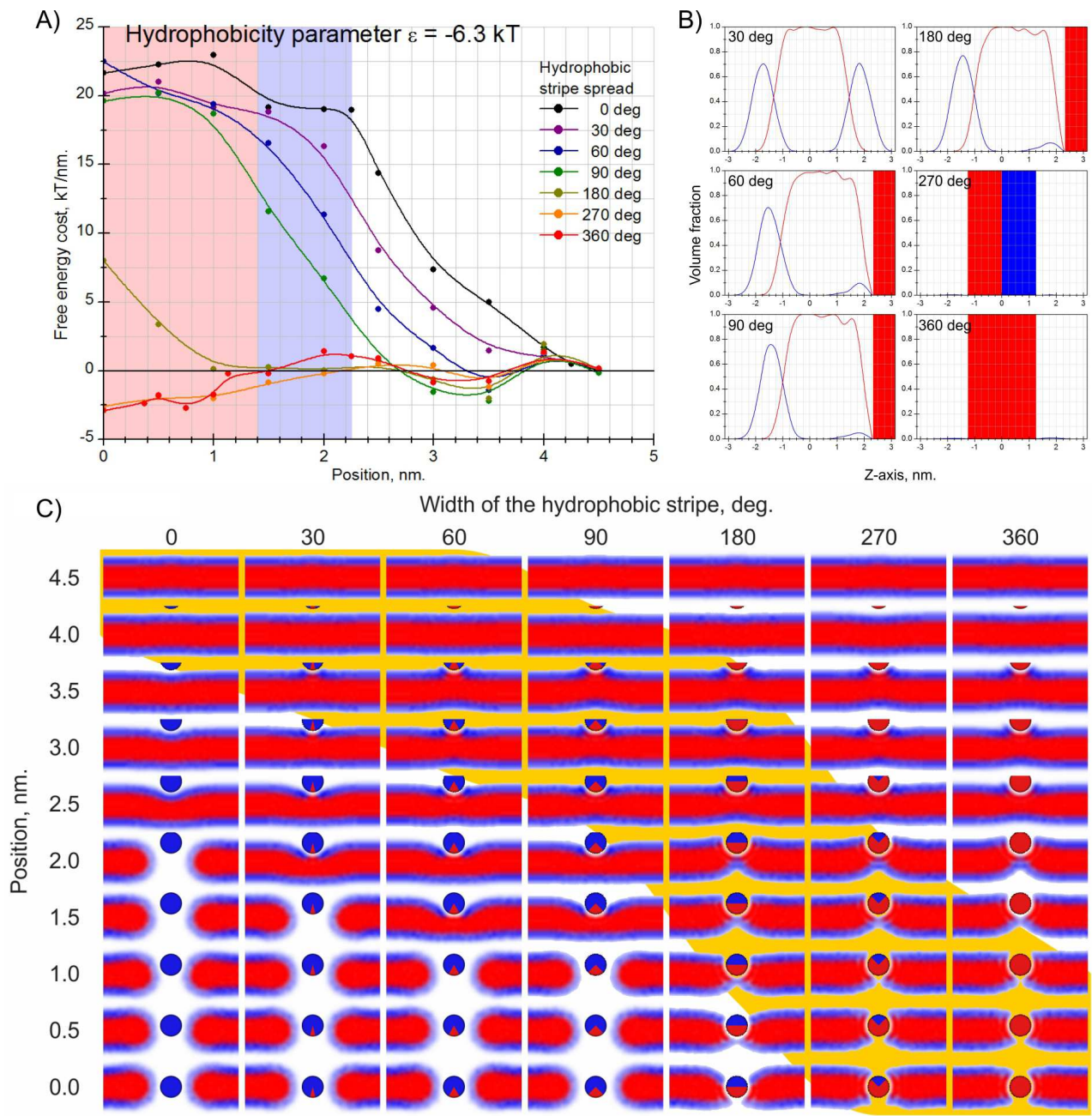}
\caption{The same as Figure \ref{Results10}, but the stripe
is less hydrophobic, $\varepsilon=-6.3$ kT.} \label{Results6}
\end{figure*}

Equilibrium structures of a rigid rod inside a phospholipid
bilayer are simulated within the SCMF theory\cite{Pogodin}. A
phospholipid molecule is represented at a coarse-grained level
within the three beads model\cite{Pogodin} which is found to
describe adequately the mechanical properties of the phospholipid
bilayer at equilibrium. The infinite rod lying parallel to the
bilayer represents the excluded area for the lipids and water
molecules, while the interactions between phospholipids in the
bilayer are described through the mean fields. The results of SCMF
calculations are presented in Figure \ref{Results10} and
Figure \ref{Results6}. The free energy differences are
calculated per unit cylinder length and neglect end effects. This
approximation of infinitely long cylinder is reasonable for
cylinders in which the lateral surface area is much larger than
the end-cap area. The interaction of lipids with the cylinder are
described by a single parameter $\varepsilon$ that is the energy
of interaction of phospholipid tails T with the surface of the
cylinder: $\varepsilon=0$ implies only steric repulsion, while
negative values reflect the attraction of T-beads to the surface
of the cylinder.

Insertion of a completely hydrophilic cylinder, $\varepsilon=0$,
into a bilayer is energetically unfavorable (black lines in Figure
\ref{Results10}(A) and Figure \ref{Results6}(A)). The
energy penalty increases with the length of the cylinder. The
hydrophilic cylinder compresses the bilayer at small penetrations
and breaks it at distances about $2.0 - 2.5$ nm from the center.
The insertion free energy increases monotonously with depth of
penetration and reaches a plateau of height $21.0 \pm 2.0$ kT/nm.
The rupture of the bilayer is similar to a first order transition
in the bulk. Two solutions, one corresponding to the compressed
bilayer and another to the broken bilayer, can coexist at some
distance, while for other distances one solution is stable and
another is metastable. Thus, the cylinder can follow the
metastable branch with higher free energy of compressed and intact
bilayer before abrupt rupture, accompanied with sudden drop of the
free energy. Another reason for the dispersion of the points at
the plateau are the error of calculations. Thus, it can be used to
estimate the accuracy of the calculations ($\pm 2$ kT/nm).

\begin{figure*}[tbp]
\includegraphics[width=12cm]{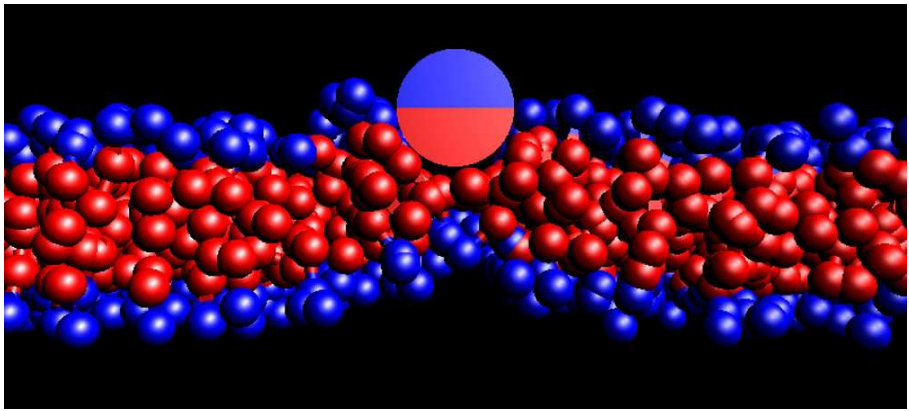}
\caption{Amphiphilic cylinder (stripe width 180 degrees,
$\varepsilon=-10.0$) and the most probable conformations of
lipids.} \label{Snapshot}
\end{figure*}

The presence of a parallel hydrophobic stripe on a hydrophilic
cylinder makes the penetration of the cylinder inside the bilayer
more favorable. This stripe is described by its width or projected
angle (in degrees) in Figure \ref{Results10} and Figure
\ref{Results6}. We explore only the orientation with the
stripe faced down to the bilayer, since this is the lowest energy
orientation, but these calculations can also be used to estimate
the energy of rotation of the cylinder in the bilayer.

For relatively narrow stripes (with the projected angle in the
range $30-90$ degrees) phospholipid heads move apart from the
cylinder to accommodate the insertion at low penetrations, which
allows for the energetically favorable contact between the stripe
and the hydrophobic core of the membrane, and leads to the wetting
of the stripe. The contact line moves up the hydrophilic cylinder
surface forming a rim and increases the contact with the bilayer
core. At higher penetrations the cylinder compresses the bilayer,
and, eventually, breaks the membrane. The breakage distance
between the cylinder and the membrane centers decreases with an
increasing width of the stripe from $1.5 - 2.0$ nm (the stripe of
$30$ degrees projected angle) to $0.0 - 0.5$ nm ($90$ degrees
stripe) and the free energy of penetration becomes lower. The
equilibrium position of the cylinder within the bilayer (yellow
band in Figure \ref{Results10}(C) corresponding to the
minimum of the energy in Figure \ref{Results10}(A)) also
shifts to the center of the bilayer with increasing width of the
stripe. Thus, the width of the stripe controls not only the
hydrophobicity of the cylinder and the insertion energy into the
bilayer, but also the equilibrium position with respect to the
bilayer core (Figure \ref{Results10}(B)).

The mean field concentration profiles in Figure
\ref{Results10}(B) and Figure \ref{Results10}(C)
demonstrate that hydrophobic interactions of $\varepsilon=-10.0$
kT are high enough for the cylinder to partially adsorb
phospholipids from the bilayer. The hydrophobic stripe is covered
with the phospholipids even after the rupture of the bilayer. For
larger widths of the hydrophobic stripe ($180-360$ degrees), the
rupture of both leaflets of the bilayer does not occur and the
cylinder is covered by phospholipids in the entire range of
penetrations from $0.0$ to $4.5$ nm from the bilayer center. The
bilayer only bends around the cylinder at small penetrations
$1.5-4.0$ nm, and covers the whole hydrophobic part of its surface
at larger penetrations $0.0-1.5$ nm.

The SCMF method provides the probability of conformations of
molecules in a given field. Thus, it allows for visualization of
the solutions of the mean field equations in the form of the most
probable conformations of molecules in the mean fields that
correspond to the solution of the equations\cite{Pogodin}. The
example of most probable conformations of phospholipids which
correspond to the equilibrium position of the amphiphilic cylinder
in the bilayer is shown in Figure \ref{Snapshot}.

At lower hydrophobicity ($\varepsilon = -6.3$ kT, Figure
\ref{Results6}), the behavior of the system is different in
many aspects. The compression of the bilayer for relatively small
($30-90$ degrees) widths of the hydrophobic stripe leads to the
rupture of the membrane farther from the bilayer center than for
the same stripes with higher hydrophobicity. Similarly, the
equilibrium position of the cylinder with the same stripes but
lower hydrophobicity is shifted closer to the surface of the
bilayer. The overall surface of narrow stripes is too small to
allow for the contact with two edges of the broken layer.

Hydrophobic stripes of larger width and $\varepsilon=-6.3$ kT
induce small perturbations to the structure of the bilayer due to
rearrangements of lipids around the stripe at relatively small
penetrations. Phospholipid heads pushed apart by the inserted
cylinder enable contact of the stripe with the hydrophobic core of
the bilayer. Penetration of the cylinder to $2.0-2.5$ nm from the
bilayer center, induces the rupture of the bilayer into two parts.
Each part of the broken layer contacts with the stripe at their
edges. The equilibrium position of the cylinders with large
hydrophobic stripes ($180-360$ degrees) is in the center of the
bilayer. However, the interaction energy $\varepsilon=-6.3$ is
insufficient for phospholipids to cover the cylinder from all
sides and the cylinder forms a pore in the bilayer (the central
cross-section concentration profiles in Figure
\ref{Results6}(B) are different from the corresponding
profiles in Figure \ref{Results10}(B) for $180-360$ degrees
stripes).

\begin{figure}[tbp]
\includegraphics{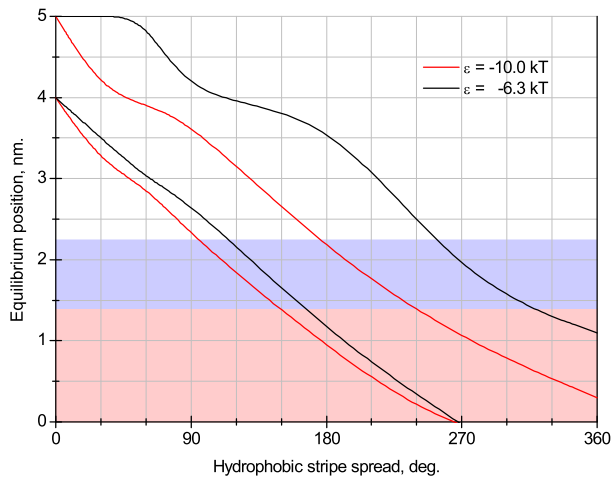}
\caption{Range of the cylinder equilibrium position \textit{versus} width
of the hydrophobic stripe on its surface. The equilibrium position
is in the region between the two red lines in the case of the
stripe with higher hydrophobicity, and between the two black lines
in the case of the stripe with lower hydrophobicity. Rose and blue
stripes designate the core and the heads regions of the bilayer.}
\label{EqAreas}
\end{figure}

The main difference between $\varepsilon=-10.0$ kT and
$\varepsilon=-6.3$ kT cases is the difference in the free energy
of insertion of the cylinder with large widths of the hydrophobic
stripe ($180-360$ degrees) (\ref{Results10}(A) and
\ref{Results6}(A)). The more hydrophobic cylinder
($\varepsilon=-10.0$ kT) is fully covered with lipids tails and
thus its free energy of insertion is much lower than that of the
partially covered cylinder with $\varepsilon=-6.3$ kT.
Furthermore, the free energy plots for more hydrophobic cylinder
$\varepsilon=-10.0$ kT have pronounced minima, while the plots
corresponding to $\varepsilon=-6.3$ kT have wide regions of
relatively constant energies around their minima which correspond
to the equilibrium position. Taking into account that the free
energy calculations presented here have an accuracy of about $\pm
2.0$ kT/nm and the minima are shallow, one can plot the areas
where the free energy cost of the cylinder insertion inside the
membrane differs from its minimum value by less than $2.0$ kT/nm
(Figure \ref{EqAreas}). The region of equilibrium positions
is wider and is on average shifted from the center of the bilayer
for lower hydrophobicity. Thus, our results suggest that the
cylinders with wide and low hydrophobicity stripes are able to
change their position within a large range around the core of the
bilayer with little or no energy cost.

The shallow minima of less hydrophobic ($\varepsilon=-6.3$ kT) and
wide ($180-360$ degrees) stripes can also be due to a
particularity of the methodology used for the calculations. The
vertical size of the simulation box is fixed for both cases and
the walls restrict the movement of the bilayer, and so the absence
of complete covering of the hydrophobic part of the cylinder with
the phospholipids in case of $\varepsilon=-6.3$ kT leaves more
free space inside the simulation box to bend the bilayer. As a
result, small shifts of the cylinder inside the box for less than
$1.5$ nm from the equilibrium, lead to bending of the bilayer
together with the cylinder without reorganization of the bilayer
structure. In contrast, there is almost no free space in the case
of the cylinder with $\varepsilon=-10.0$ kT and small deviations
of the position lead to significant reorganization of lipids
covering the hydrophobic stripe, which has a higher free energy
cost. However, this difference is only significant for two points
of the free energy corresponding to wide stripes.

Although the minimum free energy orientation is with the stripe
faced down to the bilayer, an estimate can be made of the free
energy of the cylinder rotating around its axis. In the cases of
completely hydrophilic, and completely hydrophobic cylinders,
there is no changes to the free energy. Rotation of the cylinder
with the stripe, leads to contact of the bilayer edges and the
hydrophobic core of the cylinder as well as contact between the
heads of the bilayer phospholipids with a hydrophilic part of the
cylinder. Thus, there is an energy barrier for rotation, which can
be estimated as follows. The maximum height of the barrier for
rotation of cylinder at position $0.0$ nm from the bilayer center
for $90$ degrees will be equal, approximately, to the average of
the free-energy costs of insertion into this position of
completely hydrophobic and completely hydrophilic cylinders (from
which we can estimate the energy of rotation for each position).
In all others cases the equilibrium positions found for the
orientation \textquotedblleft stripe-down\textquotedblright\ will
be true equilibrium position for any rotation of the cylinder,
because in all these positions the hydrophobic core of the bilayer
has the maximum possible contact with the hydrophobic stripe of
the cylinder. Hence, any rotation of the cylinder will not
increase the contact area, but will significantly deform the
bilayer, causing the free energy loss.

\section{Conclusion}

Our results demonstrate physical principles of the
association of rod-shaped objects with a specific surface pattern
and self-assembled phospholipid bilayers. In particular, a
hydrophobic stripe along the axis of the cylinder can mimic
self-assembled peptides with a secondary periodic structure such
as $\alpha$-helices. The position of the cylinder in the bilayer
as well as the free energy of association depend on the geometry
of the stripe, \textit{i.e.} the width and the hydrophobicity.
This microscopic information, in turn, can be used for design of
artificial polymers with enhanced association with phospholipid
bilayers.

\section{Materials and methods}

Within the SCMF theory\cite{Pogodin} the phospholipid molecule is
modeled at a coarse-grained level, while the interactions between
the molecules in the bilayer are represented by the mean fields.
The three-beads model\cite{Pogodin,Pogodin2,Pogodin3,Pogodin4}
describes a phospholipid molecule as a sequence of three spherical
beads of a radius $4.05$ \AA, one hydrophilic (H) and two
hydrophobic (T), joined consequently with the bond length $10.0$
\AA. Two T-beads of neighboring molecules interact with the energy
$\varepsilon _{TT}=-2.10$ kT if the distance between the centers
of the molecules is smaller than $12.15$ \AA. The H-beads interact
with implicit solvent molecules with the energy $\varepsilon
_{HS}=-0.15$ kT if the distance between the centers of the
molecules is smaller than $12.15$ \AA. The solvent molecules are
considered as spherical beads of the same radius $4.05$ \AA.  The
phospholipid molecule is free to bend around the central bead.

A cylinder of diameter $24.3$ \AA\ and longitudinal hydrophobic
stripe on the surface is inserted into phospholipid bilayer at
different positions in the simulation box. The position of the
cylinder is fixed parallel to the bilayer with the orientation
\textquotedblleft stripe-down\textquotedblright. The hydrophobic
part of the surface of the cylinder interacts with the T-beads
with the energy $\varepsilon $, if the centers of the molecules
are at the distance closer than $8.1$ \AA, while the hydrophilic
stripe has only steric repulsion interactions with phospholipids
and water molecules. This allows to describe the hydrophobicity of
the cylinder with a stripe by its width and a single interaction
parameter, $\varepsilon $. We consider an infinite cylinder, where
the cylinder length is equal to the size of the simulation box in
the corresponding direction, and use the periodic boundary
conditions.

In cases the cylinder penetration is unfavorable, the phospholipid
bilayer tends to bend and escape the inserted cylinder to minimize
the equilibrium free energy. To restrict the bending of the
phospholipid bilayer, we introduce hard walls at the top and
bottom of the simulation box. This allows to estimate the free
energy of insertion at a given position according to procedure
described in Ref.\citenum{Pogodin}.

\acknowledgments{PS and VAB acknowledge Spanish Ministry of education MICINN
project CTQ2008-06469/PPQ. PS, VAB and NKHS thanks the Royal
Society for the provision of an International Joint Project grant
that facilitated aspects of the work.}




\providecommand*\mcitethebibliography{\thebibliography}
\csname @ifundefined\endcsname{endmcitethebibliography}
  {\let\endmcitethebibliography\endthebibliography}{}

\end{document}